\documentclass[journal = ancac3, layout = traditional]{achemso} 

\setkeys{acs}{usetitle=true}
\setkeys{acs}{maxauthors=20}

\usepackage{color}
\usepackage{xcolor}
\usepackage{graphicx}
\usepackage{mathrsfs}
\usepackage{bm}
\usepackage{soul}
\usepackage{amssymb}
\usepackage{amsmath}
\usepackage{physics}
\usepackage{esvect}

\title[Organic Charge-Transfer Complex]{Electronic Characterization of a Charge-Transfer Complex Monolayer on Graphene}

\author{Avijit Kumar}
\email{avijitkumar@iitbbs.ac.in}
\affiliation{School of Basic Sciences, Indian Institute of Technology Bhubaneswar, Jatni, 752050 Khurda, India}
\alsoaffiliation{Department of Applied Physics, Aalto University, FI-00076 Aalto, Finland}
\author{Kaustuv Banerjee}
\affiliation{Department of Applied Physics, Aalto University, FI-00076 Aalto, Finland}
\author{Mikko M. Ervasti}
\affiliation{Department of Applied Physics, Aalto University, FI-00076 Aalto, Finland}
\author{Shawulienu Kezilebieke}
\affiliation{Department of Applied Physics, Aalto University, FI-00076 Aalto, Finland}
\author{Marc Dvorak}
\affiliation{Department of Applied Physics, Aalto University, FI-00076 Aalto, Finland}
\author{Patrick Rinke}
\affiliation{Department of Applied Physics, Aalto University, FI-00076 Aalto, Finland}
\author{Ari Harju}
\affiliation{Department of Applied Physics, Aalto University, FI-00076 Aalto, Finland}
\alsoaffiliation{Varian Medical Systems Finland, FI-00270 Helsinki, Finland}
\author{Peter Liljeroth}
\email{peter.liljeroth@aalto.fi}
\affiliation{Department of Applied Physics, Aalto University, FI-00076 Aalto, Finland}

\keywords{scanning tunneling microscopy (STM), charge-transfer complex, F$_4$TCNQ, TTF,  epitaxial graphene, charge density wave (CDW)}

\begin{document}

\begin{abstract}

Organic charge-transfer complexes (CTCs) formed by strong electron acceptor and strong electron donor molecules are known to exhibit exotic effects such as superconductivity and charge density waves. We present a low-temperature scanning tunneling microscopy and spectroscopy (LT-STM/STS) study of a two-dimensional (2D) monolayer CTC of tetrathiafulvalene (TTF) and fluorinated tetracyanoquinodimethane (F$_4$TCNQ), self-assembled on the surface of oxygen-intercalated epitaxial graphene on Ir(111) (G/O/Ir(111)). We confirm the formation of the charge-transfer complex by d$I$/d$V$ spectroscopy and direct imaging of the singly-occupied molecular orbitals. High-resolution spectroscopy reveals a gap at zero bias, suggesting the formation of a correlated ground state at low temperatures. These results point to the possibility to realize and study correlated ground states in charge-transfer complex monolayers on weakly interacting surfaces.
\end{abstract}

\maketitle


Organic charge-transfer complexes (CTCs) formed by electron-donor and -acceptor molecules are an intriguing and broad class of materials that can exhibit phenomena related to strong electron correlations and electron-phonon coupling such as 
charge and spin density waves, Mott metal-insulator transitions, charge ordering, spin-liquid phases, and superconductivity.\cite{Jerome2004review,Enoki2004review,Seo2004review,Powell2006review,Clay2018RepProgPhys,Zhang2017_AccChemRes} In bulk CTC crystals, donor and acceptor molecules typically stack in rows that maximize $\pi-\pi$ electronic overlap along the rows only.\cite{Sing2003} This anisotropy in the overlap results in pseudo one-dimensional electronic dispersion, providing a suitable platform to investigate low-dimensional, as well as low-energy, physics. 
Despite the broad spectrum of intriguing physical phenomena that have been reported in bulk CTCs, their two-dimensional (2D) films have been much less studied.\cite{Gonzalez-Lakunza2008,Fernandez-Torrente2008,Jackel2008,Clark2010,Rojas2013CTC,Jeon2016,Rodriguez-Fernandez2017,Hassanien2017,PhysRevB.81.155403} In particular, the studies have been confined to metal substrates, which strongly interact with the molecular layer and mask the intrinsic electronic properties of the CTCs.

The CTC formed out of tetrathiafulvalene (TTF) and tetracyanoquinodimethane (TCNQ) molecules is an archetypal example of a CTC. It possesses the highest bulk conductivity reported so far in a CTC and has been studied in detail.\cite{Jerome2004review,Nishiguchi1998CDW,Wang2003TTFTCNQ,Sing2003,PhysRevB.81.155403} Another widely studied system is formed by the Bechgaard salts consisting of small, planar organic molecules acting as an electron donor combined with an electron accepting small inorganic molecule. These materials are one of the most prominent examples of organic superconductors.\cite{Jerome2004review,Clark2010}

The properties of 2D films of these CTCs on metallic substrates can be strongly influenced by the underlying substrate. 
For example, it is possible to form films with other than 1:1 stoichiometry.\cite{Rojas2013CTC,Jeon2016,Rodriguez-Fernandez2017} In some cases, the effect of the substrate can be limited to doping of the film, \textit{e.g.}~in the case of the organic superconductor BETS$_2$GaCl$_4$ monolayer on Ag(111).\cite{Clark2010,Hassanien2017} On the other hand, the substrate interaction can completely dominate the low-energy electronic properties. On Au(111), TTF-TCNQ molecular states of the CTC hybridize with the metal states to form dispersive interface states.\cite{Gonzalez-Lakunza2008} Further, the unpaired electron of TCNQ molecules on the Au(111) surface exhibits the many-body Kondo effect due to screening by the substrate conduction electrons.\cite{Fernandez-Torrente2008} 
Thus, the electronic properties of a CTC, especially close to the Fermi energy, can be strongly perturbed by the metal substrate, prohibiting the study of intrinsic electronic properties of CTC. Therefore, preparing 2D films of CTCs on weakly interacting substrates is extremely desirable. Epitaxial graphene grown on Ir(111) has been shown to decouple the adsorbate layer from the underlying metal substrate allowing investigation of intrinsic electronic properties of the adsorbate layers.\cite{Kumar2017review, Kumar2018} 

Here, we a present low-temperature scanning tunneling microscopy (LT-STM) study of a 2D CTC of TTF and fluorinated TCNQ (F$_4$TCNQ) self-assembled on the surface of oxygen-intercalated epitaxial graphene on Ir(111) (G/O/Ir(111)). Sequential deposition of the molecules on this surface leads to the formation of rotationally identical domains of CTCs with alternating rows of TTF and F$_4$TCNQ lying parallel to the surface. The frontier molecular orbitals of the molecular species in the CTC, as found from scanning tunneling spectroscopy (STS), indicate charge transfer between TTF and F$_4$TCNQ molecules. High-resolution tunneling spectra exhibit a dip at Fermi Fermi energy closing at a temperature of 20 K that may be attributed to the formation of a correlated ground state in the CTC monolayer. 


\section*{Results and Discussion}

\begin{figure}[h!]
  \includegraphics[width=0.8\textwidth]{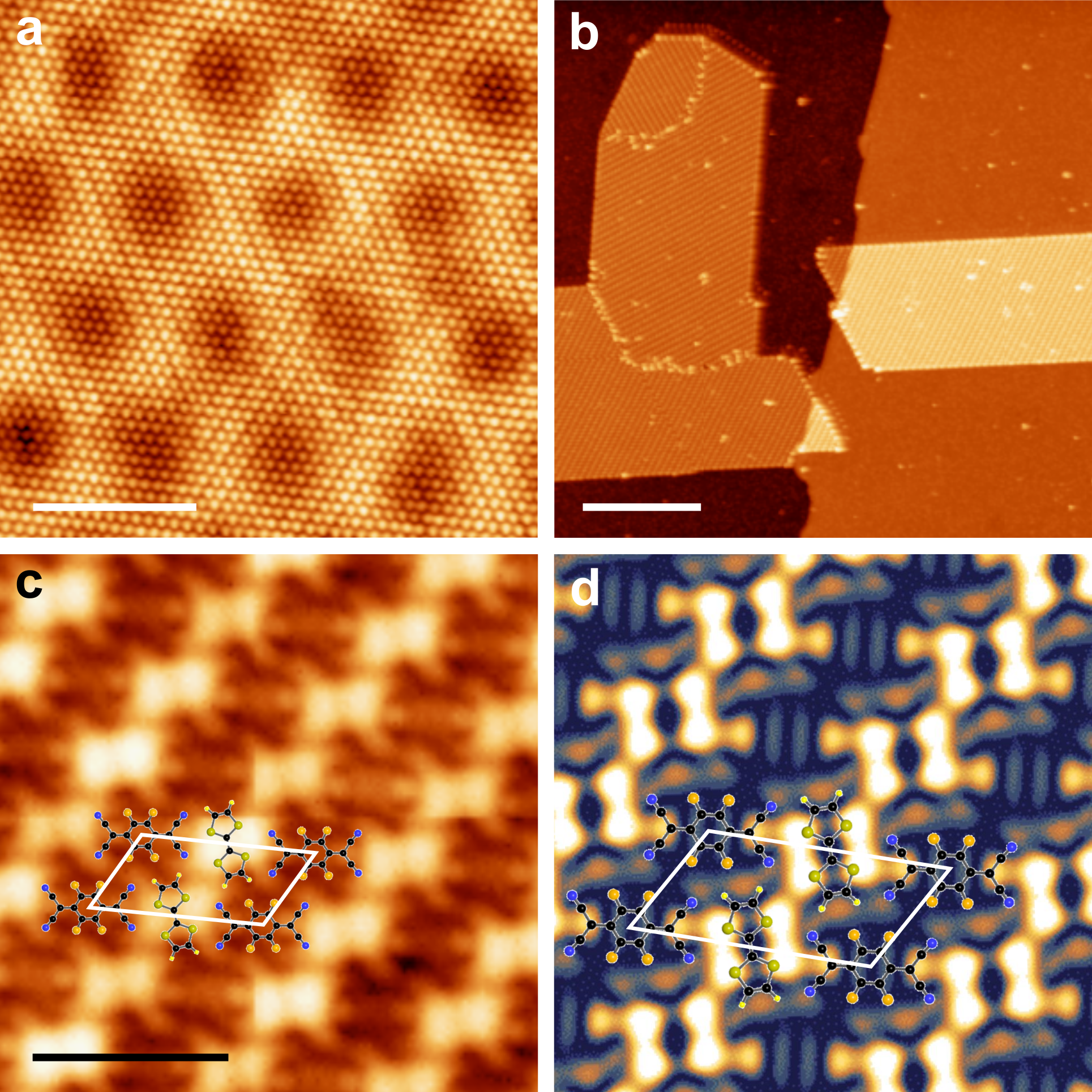}
  \caption{Assembly and structure of the CTC on oxygen-intercalated graphene. (a) STM topography image of oxygen intercalated graphene on Ir(111). The additional superstructure apart from the moir\'e is due to reconstruction of subsurface oxygen. Scale bar is 3 nm. Imaging parameters: 1.2 nA and 10 mV. (b) Few large islands of CTC on the G/O/Ir(111) surface showing various domains and the domain boundaries. Scale bar is 30 nm. Imaging parameters: 0.4 pA and 0.75 V. (c) A zoomed-in STM image of the CTC shows the arrangement of TTF and F$_4$TCNQ molecules. Each molecule forms a row next to the row of the other molecule. A molecular structure along with a unit cell is overlaid to elucidate the molecular arrangement within the unit cell. Scale bar is 2 nm. Imaging parameters: $\sim$5 pA and 0.1 V. (d) A DFT simulated STM image of the CTC close to the Fermi energy resembles the recorded topography closely. Molecular structure and unit cells are overlaid for clarity.}
  \label{fig:fig1}
\end{figure}

Figure~\ref{fig:fig1} describes the assembly and structure of the TTF-F$_4$TCNQ CTC on a G/O/Ir(111) surface. The sample preparation is described in detail in the Methods section. Briefly, we grow a near monolayer coverage of graphene on Ir(111) by a combination of temperature programmed growth (TPG) and chemical vapour deposition (CVD), as described previously,\cite{NDiaye2008,coraux2009growth,Hamalainen2013} followed by oxygen intercalation to electronically decouple graphene from the underlying substrate.\cite{Martinez-Galera2016} Finally, the molecules are deposited at low temperatures ($\approx 100$ K), followed by annealing at room temperature for 15-45 mins to allow the formation of highly ordered CTC islands.

Figure~\ref{fig:fig1}a shows an STM topography image of oxygen intercalated graphene on Ir(111). The surface contains the periodic moir\'e pattern of a G/Ir(111) surface with a periodicity of 25.4 \r{A}. 
The additional superstructure visible on the surface is due to patches of ($2\times1$) reconstruction of subsurface oxygen which is consistent with an earlier report.\cite{Martinez-Galera2016} Oxygen intercalation leads to  decoupling of graphene from Ir, which is indicated by the short-range d$I$/d$V$-spectroscopy of the surface showing a phonon gap of $\sim$160 mV \cite{Zhang2008,Halle2018_NanoLett} (see Supporting Information (SI) Fig.~S1a). Oxygen intercalation also results in strong \textit{p}-doping of graphene by $\sim$0.5 eV,\cite{Ulstrup2014} which increases the work function to $\sim$5.1 eV. This can be independently verified by measuring d$I$/d$V$ spectra at high bias with the feedback loop on $-$ here, the field-emission resonances allow to estimate the substrate work function \cite{Binnig1985_prl,Lin2007,Schulz2013,Schulz2014_PRB} (See SI Fig.~S1b). 

Figure~\ref{fig:fig1}b shows an STM topograph of large islands of ordered CTC assembled on a G/O/Ir(111) surface. The long-range ordering is the result of the post-deposition room-temperature annealing; directly after the low-temperature deposition, we observe disordered islands on the surface (see SI Fig.~S2). The CTC islands grow across the step edges in carpet-like fashion \cite{banerjee2016flexible,Yan2020_CuDCA} and contain various domains rotated with respect to each other. Analysis of several images reveals a total of six domain orientations rotated w.r.t.~each other in multiples of 30$^\circ$. Figure~\ref{fig:fig1}c shows a zoomed-in STM image to identify arrangement of TTF and F$_4$TCNQ molecules within the CTC islands. As evident from the STM image, there are two different rows of molecules: one is composed of TTF and the other of F$_4$TCNQ molecules. Rows of TTF and F$_4$TCNQ are lying alternately on the surface. The molecular structure obtained from density functional theory (DFT) calculations (see below) has been overlaid on the STM image for clarity. The molecular rows are found to be at an angle of $\pm$12$^\circ$ w.r.t.~graphene's zigzag direction for each domain. The unit cell of the CTC is shown by a parallelogram with lattice parameters \textit{a} = 18.5 ($\pm$0.5) \r{A}, \textit{b} = 9.5 ($\pm$0.5) \r{A}, $\theta$ = 56 ($\pm$2)$^\circ$. This is the most common phase we observe for this stoichiometry ((F$_4$TCNQ$)_1$(TTF)$_1$) of the molecules. At a slightly different stoichiometry ((F$_4$TCNQ)$_x$(TTF)$_{1-x}$), we have observed a checkerboard phase of the CTC where only F$_4$TCNQ rows are present and TTF molecules are dispersed across in a checkerboard fashion (see SI Fig.~S3).

In order to further elucidate the structure of the molecular layer, we carried out a broad structural search for different possible geometries using DFT (see Methods for details). We performed full structural relaxations of 300 CTC monolayers sampled by varying intermolecular distance, bond angles, and alignment with respect to the underlying graphene. The initial structures are systematically generated but done ``by hand" without any input from machine learning or structure search algorithms.\cite{Egger/etal:2020,Jarvi/Rinke/Todorovic:2020} After relaxation, the structures are sorted by formation energy. One of the low energy conformations closely matches the experimental structure both in terms of the unit cell dimensions (\textit{a} = 17.78 \r{A}, \textit{b} = 8.89 \r{A}, $\theta$ = 60$^\circ$) and the relative orientation w.r.t.~the graphene lattice (13.89$^\circ$). A DFT simulated STM image (at Fermi energy) is shown in Fig.~\ref{fig:fig1}d for the optimized geometry; it closely resembles the STM image shown in Fig.~\ref{fig:fig1}c. 

We have also looked at the assembly of single component F$_4$TCNQ and TTF layers on the G/O/Ir(111) surface. A sub-monolayer coverage of F$_4$TCNQ molecules forms chain-like structures (in contrast to non-planar adsorption on the G/Ir(111) surface \cite{Kumar2017}). On the other hand, TTF molecules tend to assemble in a close-packed geometry on the G/O/Ir(111) surface. The assembly of F$_4$TCNQ and TTF molecules is shown in SI Figs.~S4 and S5. 

\begin{figure}[h]
  \includegraphics[width=0.9\textwidth]{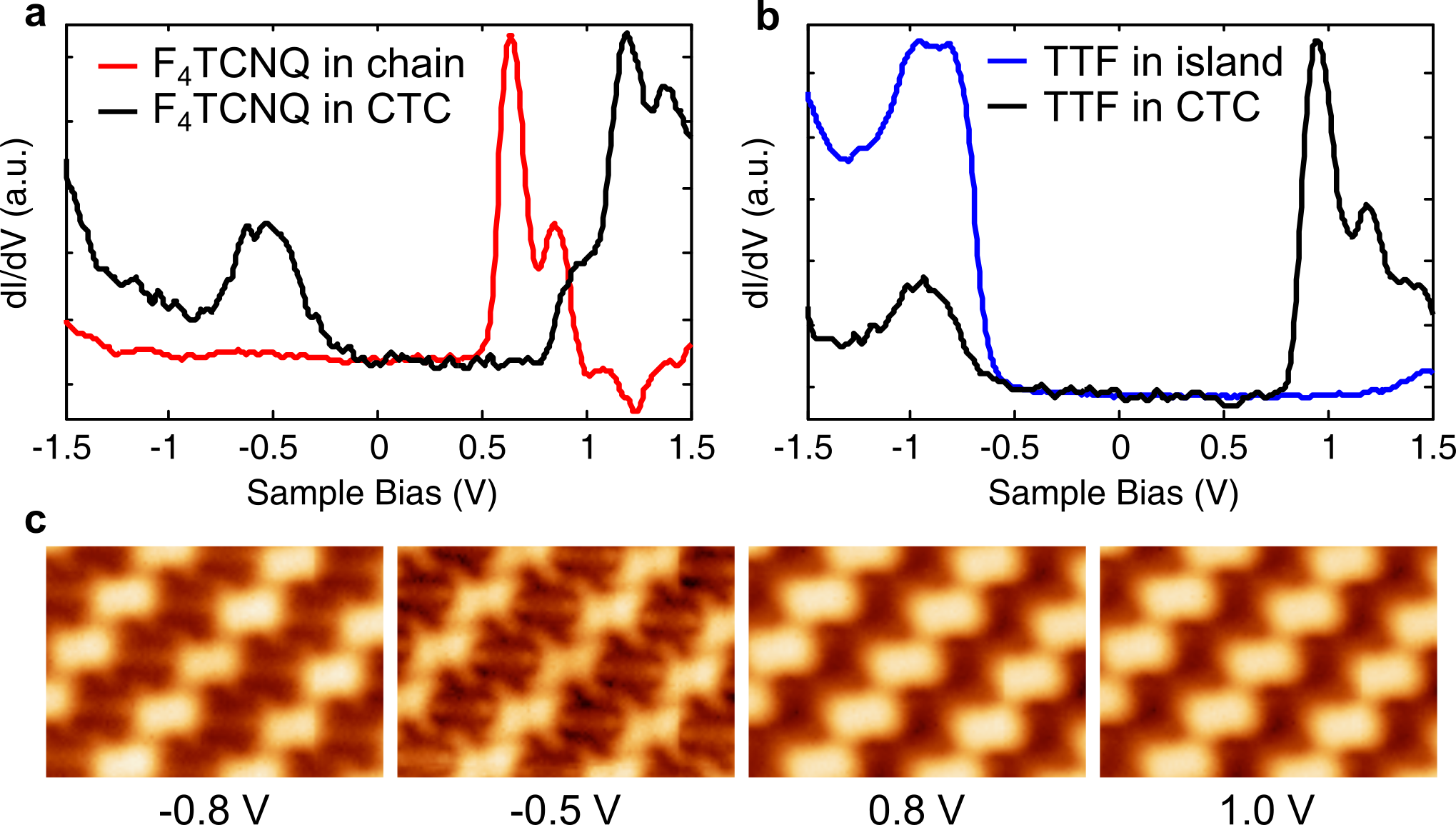}
  \caption{Charge transfer across the molecules. (a) Long range d$I$/d$V$-spectra on F$_4$TCNQ molecules in a single-component chain on the G/O/Ir(111) surface (red line) and on the F$_4$TCNQ sites in the CTC (black line). (b) Long range d$I$/d$V$-spectra on TTF molecules in a single component assembly on G/O/Ir(111) (blue line) and on the TTF sites in the CTC (black line). (c) Bias dependent STM images of the CTC at the sample biases indicated in the figure. Size of each image is $4.7\times3.2$ nm$^2$.} 
  \label{fig:fig2}
\end{figure}

Figure \ref{fig:fig2} shows the experimental verification of charge transfer between TTF and F$_4$TCNQ molecules in the CTC by d$I$/d$V$ spectroscopy and STM imaging. Fig.~\ref{fig:fig2}a compares long-range d$I$/d$V$ spectra recorded on F$_4$TCNQ molecules in single component chains to those recorded in the CTC. The spectrum on the molecule in the chain shows a resonance corresponding to the lowest unoccupied molecular orbital (LUMO) at 0.64 V without any features at negative bias. This indicates that the F$_4$TCNQ molecules on G/O/Ir(111) are neutral, in contrast to F$_4$TCNQ molecules on a G/Ir(111) surface, where they are charged at lower sites of the moir\'e pattern.\cite{Kumar2017} This difference is likely due to the increased work function of graphene due to oxygen intercalation. The spectrum recorded on a F$_4$TCNQ molecule in the CTC, on the other hand, shows two peaks at -0.44 V and 1.2 V. 
Fig.~\ref{fig:fig2}b compares the d$I$/d$V$ spectrum on TTF molecules from the pristine assembly on a G/O/Ir(111) surface to that of TTF molecules from the CTC. Here, d$I$/d$V$ spectrum on TTF molecule shows a peak at -0.8 V, corresponding to the highest occupied molecular orbital (HOMO) of a neutral TTF molecule. Despite the high work function of the surface ($\sim$5.0 eV), the TTF molecules stay neutral. In the CTC, the spectrum on TTF molecules shows two peaks at -0.9 V and 0.95 V (similar to the two peaks on an F$_4$TCNQ molecule). The assignment of these peaks is done on the basis of images recorded at sample biases at -0.5 and 0.8 V. The image at 0.8 V shows a relatively prominent TTF HOMO, while the image at -0.5 V shows a relatively prominent F$_4$TCNQ LUMO \cite{Kumar2017} (see Fig.~2c). Electron transfer from donor TTF to acceptor F$_4$TCNQ molecules results in splitting of the TTF HOMO (-0.8 eV peak) into singly occupied (SOMO, -0.95 V peak) and singly unoccupied molecular orbitals (SUMO, 0.95 V peak). Similarly, the F$_4$TCNQ LUMO (0.64 V peak) splits into SOMO (-0.44 V peak) and SUMO (1.2 V peak) after accepting an electron. Consequently, the TTF molecule acquires a positive charge while F$_4$TCNQ molecules become negatively charged in the CTC. The charge transfer between the molecules is also supported by DFT calculations, and based on Hirshfeld charge analysis \cite{hirshfeld} it amounts to $\sim$0.55 e in this configuration. Each N atom gains $\sim$0.2 e and redistribution of the remaining charge makes up the difference. The calculated band structure of the monolayer CTC (Fig.~\ref{fig:fig1}d) is shown in SI Fig.~S6. From the band structure, it is evident that there is also a charge transfer from graphene to the CTC monolayer and a finite electronic coupling in the molecules along certain directions of reciprocal space ($\Gamma$-K and $\Gamma$-Y). However, the bandwidth is relatively small ($\sim100$ meV), indicating that the coupling is quite weak.

\begin{figure}[h]
  \includegraphics[width=0.85\textwidth]{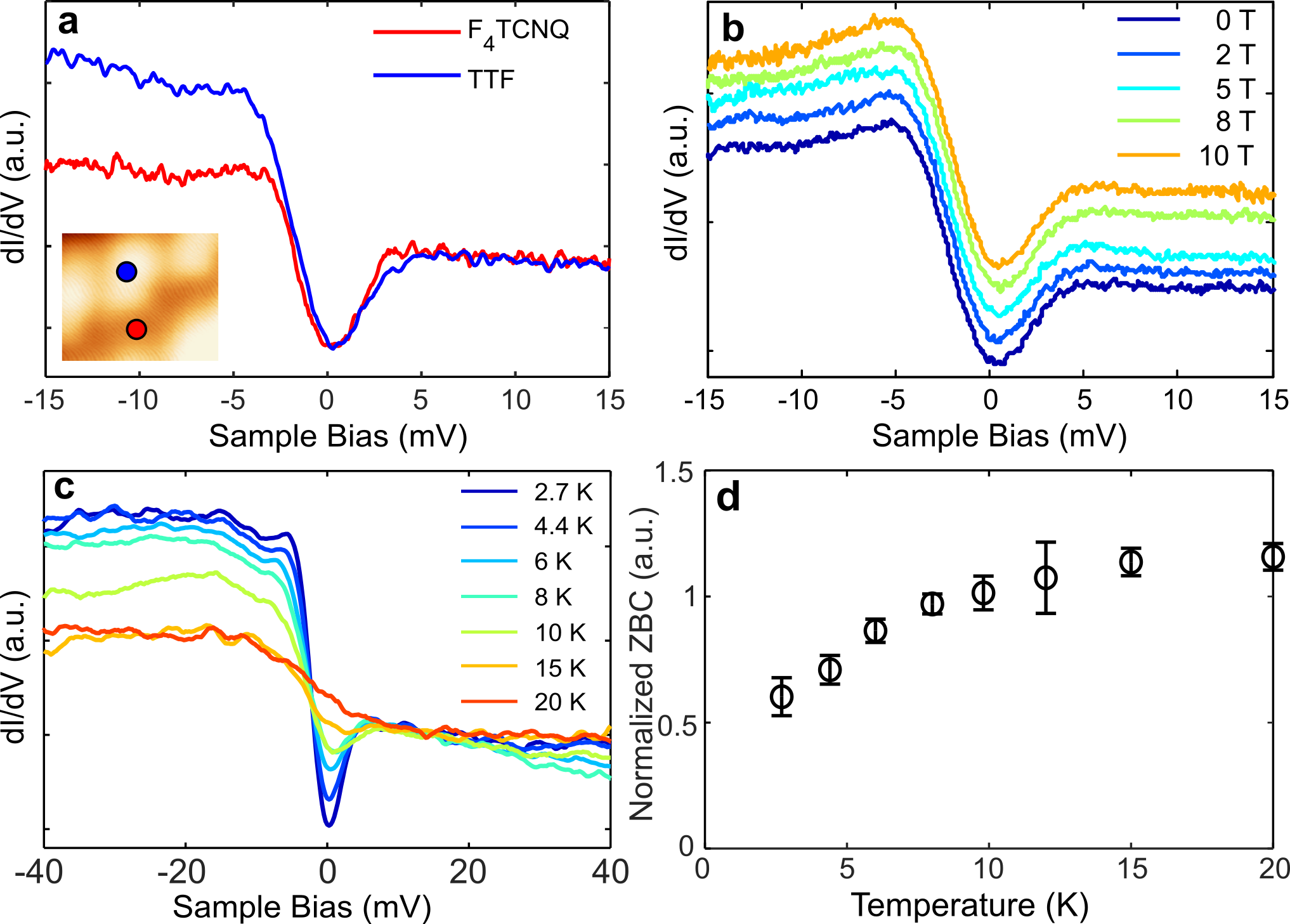}
  \caption{Short-range d$I$/d$V$-spectroscopy on the CTC. (a) Short-range d$I$/d$V$-spectroscopy on the TTF and F$_4$TCNQ sites in the CTC show a dip at zero bias. (b) Magnetic field dependent d$I$/d$V$-spectra on a TTF site in the CTC shows that the shape and size of the zero-bias dip does not change with magnetic field up to 10 T. (c) Temperature-dependent d$I$/d$V$-spectra on a TTF site in the CTC show that the dip is washed away with increasing temperature and the asymmetric background is also decreased at higher temperatures. (d) Temperature dependence of the zero-bias conductance (ZBC, normalized at the d$I$/d$V$ at bias of 20 mV) shows saturation at 15-20 K.}
  \label{fig:fig3}
\end{figure}

Interestingly, high-resolution d$I$/d$V$ spectra on both molecules contains a dip close to zero bias which has pronounced asymmetry on TTF sites as is shown in Fig. \ref{fig:fig3}a. 
To investigate its origin, we have examined its dependence on temperature and on out-of-plane magnetic field. Care was taken to record these spectra on the same molecule and with the same microscopic tip apex. Fig. \ref{fig:fig3}b shows magnetic field dependent d$I$/d$V$ spectra on the TTF sites of the CTC lattice in the range of 0 to 10 T. There is no measurable change in either the shape and size of the dip, or the observed asymmetry up to magnetic field of 10 T. On the contrary, a clear temperature dependence is observed from Fig.~\ref{fig:fig3}c, which shows the temperature-dependent d$I$/d$V$-spectroscopy recorded on TTF sites of the CTC from 2.7 K to 20 K (data on the F$_4$TCNQ site is shown in the SI Fig.~S7a). The asymmetric dip is most prominent at the lowest temperature of 2.7 K. The dip amplitude decreases with increasing temperature and at 20 K only a step at zero bias remains. The temperature dependence of the zero bias conductance (ZBC) extracted from these spectra 
clearly exhibits the saturation of the ZBC at temperatures between 15-20 K. This change in the ZBC indicates the presence of a low-temperature correlated state, which we discuss in more detail below. 

\begin{figure}[h]
  \includegraphics[width=0.8\textwidth]{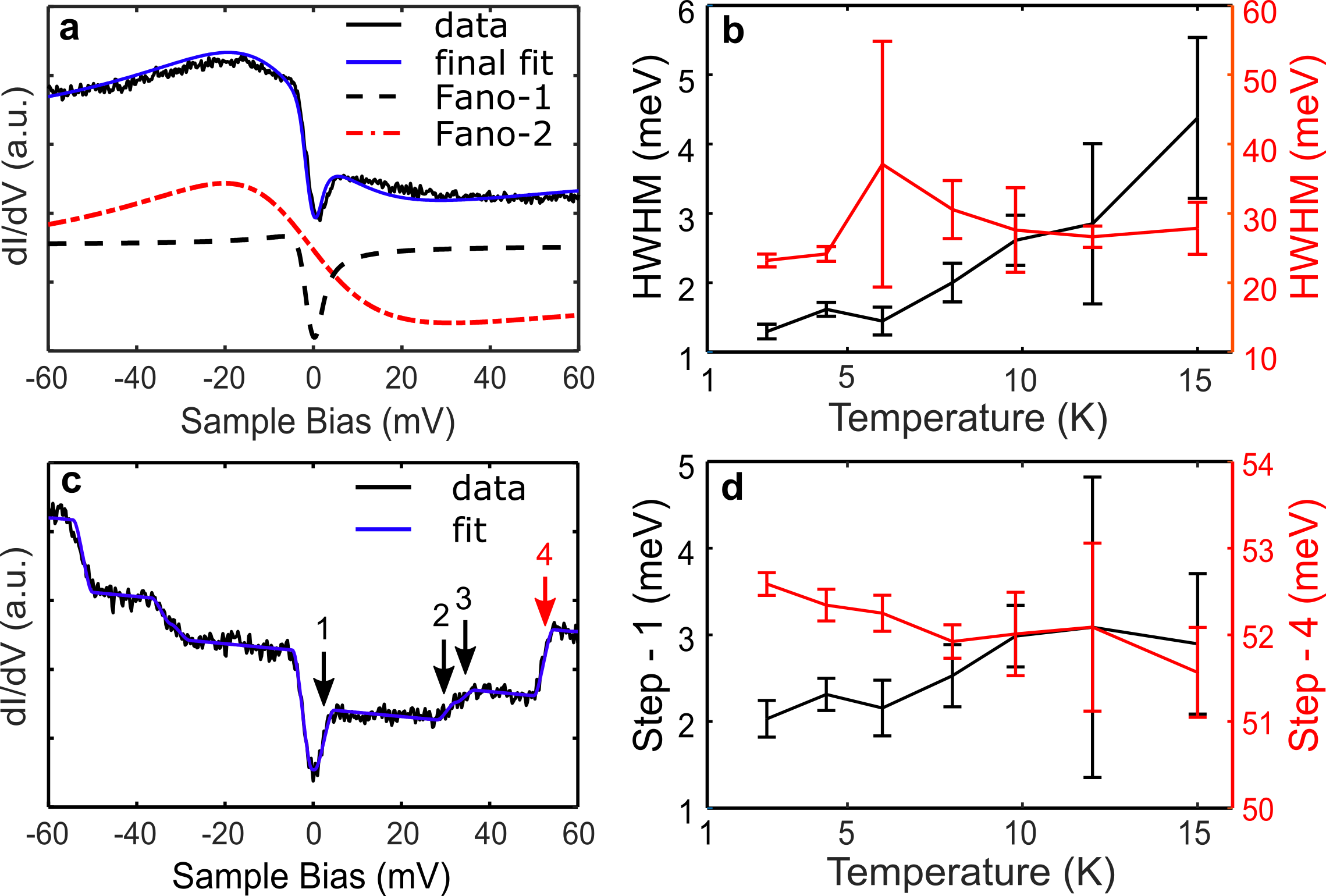}
  \caption{Deconvoluting the low-bias features of the d$I$/d$V$ spectra. (a) Short-range d$I$/d$V$-spectrum on on TTF molecules. The curve has been fitted with sum of two Fano functions: Fano-1 (broken black line) represents the central dip and Fano-2 (red line) represents the step. Final fit is indicated by blue line. (b) Temperature-dependent evolution of HWHM extracted from the two Fano function (Fano-1: left, Fano-2: right) from the fits.  (c) Short-range d$I$/d$V$-spectroscopy on CTC islands, recorded on a F$_4$TCNQ molecule showing steps at energies $\sim$2 (shown by arrow 1), $\sim$31 (arrow 2), $\sim$35 (arrow 3) and $\sim$52 meV (arrow 4). (d) Temperature-dependent evolution of the steps at $\sim$2 meV (Step-1: left) and at $\sim$52 meV (Step-4: right). }
  \label{fig:fig4}
\end{figure}

The temperature dependent spectroscopy shows that the overall asymmetry of the spectra and the amplitude of the dip reduces with increasing temperature. At 20 K, the dip feature is no longer visible while the asymmetry (a step at Fermi-energy) is still present in the spectra. This suggests that the spectrum can be deconvoluted into a dip and a step - the dip vanishes at 20 K while the step still remains visible at that temperature. 
The deconvolution of a spectrum measured on a TTF site in the CTC is shown in Fig.~\ref{fig:fig4}a. The entire spectrum (note the wider bias range here compared to Fig.~\ref{fig:fig3}a) can be well fitted (details of the fittings are described in the Methods section) by a sum of two Fano lineshapes.\cite{fano1961effects} The effect of the spectral broadening due to the bias modulation and thermal broadening have been deconvoluted (see \textit{Methods} section) to obtain the intrinsic width of the lineshapes. Fig.~\ref{fig:fig4}b summarizes the temperature dependence of the half-width half-maximum (HWHM) of the two Fano lineshapes used to fit the spectra on the TTF site. The HWHM of the Fano lineshape corresponding to the dip at zero bias (Fano-1) shows a clear scaling with temperature. On the other hand, the HWHM of the step-like Fano lineshape (Fano-2) has a weaker temperature dependence. While the Fano lineshape is taken here as a phenomenological description of the measured spectra, the choice is not completely arbitrary, as it typically arises in situations where there are two interfering tunneling pathways present. For example, it is widely observed on Kondo impurities, where the interference occurs between a direct tip-sample tunneling and tunneling path \textit{via} the Kondo impurity.\cite{li1998kondo,madhavan1998tunneling,nagaoka2002temperature,Ternes2015_review} In fact, a spectral shape combining a step-like Fano lineshape with a smaller energy gap-like feature - very similar to our measurements - has been observed on the heavy fermion compound URu$_2$Si$_2$.\cite{Aynajian10383} There, the spectral response was explained by a combination of Kondo screening of the uranium $f$-electrons and the gap-like feature resulting from a transition to a hidden order phase at low temperatures. 

Intriguingly, the d$I$/d$V$-spectra recorded on the F$_4$TCNQ molecules of the CTC (Fig.~\ref{fig:fig4}c - the bias range is again wider than in Fig.~\ref{fig:fig3}a) show additional step-like features at higher biases, \textit{viz}.~at $\pm31$, $\pm35$ and $\pm52$ mV. These steps 
can be attributed to inelastic electron tunneling processes similar to molecular vibrations of negatively charged F$_4$TCNQ molecules.\cite{Garnica2014,Fernandez-Torrente2008} 
The tunneling electrons can excite a molecular vibration once the sample bias matches the energy of the corresponding vibrational mode.\cite{Lorente2001_PRL,delaTorre2017_PRL,IETS_review} The inelastic process corresponds to opening of an additional tunneling channel and a sudden increase in the tunneling conductance. To corroborate this picture, we assess the phonon modes for the CTC monolayer with DFT (details in the Methods). There is good agreement between the energies of the measured steps and the calculated energies of certain CTC phonon modes with a high electron-phonon coupling strength. Additionally, the calculated modes with strong coupling strength near the energies of the inelastic steps are dominated by F$_4$TCNQ vibrations (see SI Fig.~S8 for details). This is consistent with our experiments, where we see the inelastic steps only on the F$_4$TCNQ sites of the CTC.

Although DFT calculations indicate presence of intermolecular phonon modes with energies of a few mV, the temperature dependence of the dip close to zero bias does not fit with thermally broadened inelastic steps. If we force a fit with an inelastic step to the data (feature marked with ``1'' in Fig.~\ref{fig:fig4}b), the position of the fitted step would be strongly temperature dependent (Fig.~\ref{fig:fig4}d, black symbols), which is not expected for inelastic features. The zero bias feature also washes out more quickly with temperature than what would be expected for a vibrational transition, and at 15 K or above, only an asymmetric step remains, supporting the notion that it is a result of a gap closing transition. This is illustrated in Fig.~S7b, which shows the expected temperature dependence for an inelastic step using the parameters extracted from the experimental spectrum acquired at $T=2.7$ K. As can be clearly seen, the predicted trend does not match the experimental results in Fig.~S7a, which gives a strong indication that the zero-bias dip feature does not correspond to inelastic steps.

Considering the width of the dip, we should be able to resolve a possible magnetic field induced splitting if this feature was arising from any spin-related phenomena such as the Kondo effect or spin-flip inelastic transitions.\cite{Ternes2015_review,Ternes2017_review} However, we do not observe any such changes with a magnetic field up to 10 T as shown in Fig.~\ref{fig:fig3}b. While the Kondo effect has earlier been observed in an TTF-TCNQ CTC monolayer on Au(111),\cite{Fernandez-Torrente2008} the Kondo coupling is expected to be generally weak on graphene.\cite{Fritz2013_RepProgPhys} Finally, experiments on CTCs deposited on graphene directly on Ir(111) show very similar response (see SI Fig.~S9). The two substrates differ significantly in terms of the doping level of graphene, which is expected to have a marked influence on the Kondo temperature.\cite{Fritz2013_RepProgPhys,Chen2011,Jiang2018} 
Further, CTCs are also known for exhibiting superconductivity. But in light of the spectroscopy measurements in high magnetic fields, superconductivity origin of the dip at Fermi energy is also very unlikely. One would expect either quenching or at least changes in the superconducting gap under high field. We also do not observe coherence peaks in the spectra that are usually associated with superconductivity.

\begin{figure}[h]
  \includegraphics[width=0.65\textwidth]{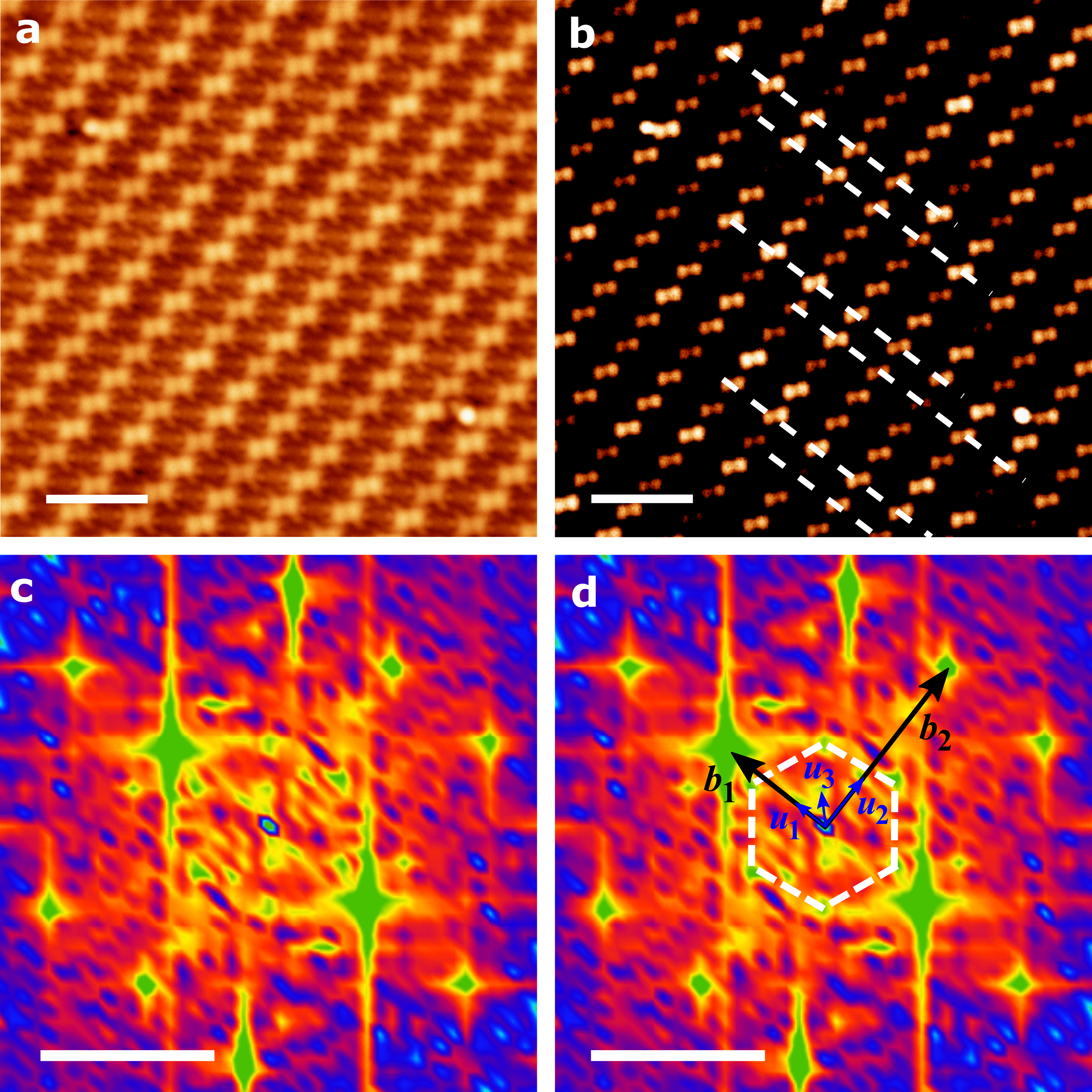}
  \caption{(a) STM topography image of CTC at imaging parameters: 5 pA and -500 mV. The scale bar is 3 nm. (b) A contrast-optimized version of the topography in panel (a) shows periodic topography modulations (white lines are guide to the eyes). (c,d) Two-dimensional fast-Fourier transform (2D-FFT) of panel 
  (a) shows features corresponding to the CTC rectangular lattice (marked by vectors \textbf{\emph{b$_{1}$}} and \textbf{\emph{b$_{2}$}}), spots due to the underlying graphene moir\'e (white hexagon), and charge density wave modulations by vectors \textbf{\emph{u$_{1}$}}, \textbf{\emph{u$_{2}$}}, and \textbf{\emph{u$_{3}$}}. CDW wavelengths corresponding to \textbf{\emph{u$_{1}$}},  \textbf{\emph{u$_{2}$}} are approximately $3.25\times l_1$ and $3.25\times l_2$, while that corresponding to \textbf{\emph{u$_{3}$}} is $\sim$5 nm. The scale bar is 1 nm$^{-1}$.
    }
  \label{fig:cdw}
\end{figure}

The remaining explanations consistent with the spectral feature and its dependence on magnetic field and temperature include the formation of a charge-density wave (CDW) or Peierls instability at low temperatures; these correlated ground states have been commonly observed in bulk CTC materials.\cite{Jerome2004review,Nishiguchi1998CDW} The structure of this compound, both in bulk and in our monolayer is anisotropic: there is much stronger electronic coupling along a certain lattice direction than in the perpendicular direction. This is also evident in the calculated band structure shown in Fig.~S6. This kind of an anisotropic bandstructure is favourable for the formation of a CDW state as it naturally provides Fermi surface nesting. This leads to the CDW driven by e-ph coupling, which is also in line with the picture for the bulk TCNQ-TTF phases.\mbox{\cite{Jerome2004review}} 
The temperature dependence of the ZBC (Fig.~\ref{fig:fig3}d) clearly indicates a transition temperature of 15-20 K, which is close to the expected temperature range of a CDW or Peierls transition; for example, in the bulk TTF-TCNQ CTC this is 54 K.\cite{PhysRevB.16.5238} Finally, the ground state associated with CDW breaks the symmetry of the system and results in a superstructure arising from modulations in electron density or CTC atomic structure.
Fig.~\mbox{\ref{fig:cdw}}a shows STM topography image of the CTC. Contrast-optimized version of the same image in Fig.~\mbox{\ref{fig:cdw}}b shows periodic modulation of the topography which can better be understood using 2D FFT. White lines are guide to the eyes. Fig.~\mbox{\ref{fig:cdw}}c and d show the 2D-FFT images of the topography with various spots identified. 
The set of spots marked by vectors \textbf{\emph{b$_{1}$}} and \textbf{\emph{b$_{2}$}} corresponds to the CTC rectangular lattice (see SI Fig.~S6) while the spots due to the underlying graphene moir\'e is indicated by a white hexagon. The set of vectors indicated by \textbf{\emph{u$_{1}$}}, \textbf{\emph{u$_{2}$}}, and \textbf{\emph{u$_{3}$}} indicate the presence of longer wavelength charge-density wave modulation. CDW wavelengths corresponding to \textbf{\emph{u$_{1}$}}, \textbf{\emph{u$_{2}$}} are approximately $3.25 \times l_1$ and $3.25\times l_2$, while that corresponding to \textbf{\emph{u$_{3}$}} is $\sim$5 nm. Here, \textbf{\emph{l$_{1}$}} and \textbf{\emph{l$_{2}$}} are real space lattice vectors perpendicular to and along the TTF/F$_4$TCNQ molecular rows, respectively. 
This provides further evidence of the presence of CDW/Peierls ground state in the TTF-F$_4$TCNQ CTC monolayer at low temperatures causing a gap in the density of states at the Fermi energy.

\section*{Conclusions}
In conclusion, we have synthesized a monolayer of charge-transfer complex TTF-F$_4$TCNQ on a weakly interacting epitaxial graphene substrate, and have investigated its intrinsic electronic properties. TTF and F$_4$TCNQ molecules assemble into close-packed islands with alternating rows of TTF and F$_4$TCNQ molecules in a 1:1 stoichiometry. Low-temperature STM and STS measurements confirm the formation of a charge-transfer complex with d$I$/d$V$ spectra consistent with the presence of TTF cations and F$_4$TCNQ anions. High-resolution spectroscopy at low-temperatures and high magnetic fields show formation of a correlated ground state related to a CDW or Peirls instability with a transition temperature of 15-20 K. This work demonstrates CTC monolayers as intriguing example of two-dimensional materials with low-temperature correlated ground states.

\section*{Methods}
\textit{Sample preparation.} The experiments were carried out in ultra-high vacuum (UHV), low-temperature scanning tunneling microscopes (STMs) (Createc LT-STM and Unisoku USM-1300). Both STMs are equipped with a preparation chamber and operate at a base pressure lower than $1\times10^{-10}$ mbar. The sample was prepared by depositing F$_4$TCNQ and TTF molecules sequentially on an oxygen-intercalated graphene on Ir(111) substrate. The Ir(111) surface was cleaned by repeated cycles of sputtering using Ne ions at energy 1.5 kV and annealing at 900 $^\circ$C in an oxygen environment, followed by flashing to 1300 $^\circ$C. Epitaxial graphene was grown using ethylene gas with a combination of temperature programmed growth (TPG) and chemical vapour deposition (CVD) steps to achieve a nearly full monolayer coverage of graphene.\cite{NDiaye2008,coraux2009growth,michely_prl_2011,Hamalainen2013} In the TPG step, the cleaned Ir(111) substrate was exposed to the ethylene gas for one minute at a pressure of $1\times10^{-6}$ mbar followed by heating the substrate to 1300 $^\circ$C. The CVD step was carried out at this temperature by exposing the substrate to ethylene gas at $3\times10^{-7}$ mbar for 60 s. This gives nearly a monolayer coverage of graphene on Ir(111) (G/Ir(111)). Oxygen intercalation of G/Ir(111) (G/O/Ir(111)) was carried out by exposure of $9\times10^4$ L oxygen at 225$^\circ$ C as reported by Ref.~\cite{Martinez-Galera2016}. 

The charge-transfer complex (CTC) was synthesized by first depositing $\sim$0.25 monolayer of F$_4$TCNQ molecules on a G/O/Ir(111) surface at low substrate temperature ($\approx 100$ K), followed by deposition of a similar amount of TTF molecules at a similar substrate temperature. This resulted in disordered islands of CTC on the surface. The sample was annealed at room temperature for 15-45 mins. to allow the formation of highly ordered CTC islands. While F$_4$TCNQ molecules were evaporated using a Knudsen cell heated to 92 $^\circ$C, TTF molecules were evaporated from a home-made evaporator kept at temperature 23 $^\circ$C. The deposited amounts of the two molecules were adjusted to 1:1 stoichiometry (each of them at less than a half monolayer coverage). Subsequently, the sample was transferred into the low-temperature STM housed within the same UHV system. 

\textit{STM measurements.} The STM experiments were carried out at a temperature of 4.2 K unless otherwise stated. Temperature-dependent measurements were carried out in the Createc STM, while magnetic field dependent measurements were carried out in the Unisoku STM. For the measurements at 2.7 K, the LHe cryostat of the STM was pumped, while measurements at temperature higher than 4.2 K was achieved by heating the STM by a Zener diode installed on the STM scanner. To avoid any ambiguity, the temperature-dependent measurements were carried out on the same F$_4$TCNQ and TTF molecules of the CTC assembly using the same tip. Similar precautions were taken for the magnetic field measurements as well, where the same molecules and the tip was used for the full-range of the magnetic field sweep.  STM measurements were carried out using mechanically cut Pt/Ir tips. d$I$/d$V$-spectroscopy was performed using a standard lock-in technique, where a voltage modulation with amplitude of 10-15 mV and 1-2 mV signal has been used for long-range and short-range spectroscopies, respectively. WSxM \cite{wsxm} and Gwyddion (\url{http://gwyddion.net/})\cite{gwyddion2} software were used to process all the STM images.

\textit{Fitting of the d$I$/d$V$-spectra.} We use two Fano lineshape functions to fit the short-range d$I$/d$V$ spectrum in Fig.~\ref{fig:fig4}a. The Fano lineshape function is: 
\begin{equation*}
f_\mathrm{Fano}(\epsilon) = A\frac{(q + \frac{\epsilon-\epsilon_0}{\Gamma})^2}{1 + (\frac{\epsilon-\epsilon_0}{\Gamma})^2} + c_1 
\end{equation*}

where \textit{A} is the prefactor, $\epsilon$ is the energy, $\epsilon_0$ is offset from zero, $\Gamma$ is the half width at half maximum, $q$ is the Fano parameter, and c$_1$ is a constant background term. We first fit the step-like Fano lineshape to capture the step of the spectrum (Fano-2) by excluding the central dip during the fitting. Further, we subtract the step-like Fano fit (red line in Fig.~\ref{fig:fig4}a) from the spectrum to get a central dip which is fitted again using a dip-like Fano lineshape (Fano-1). The fitting process is repeated for all the recorded spectra at the indicated temperatures to extract HWHM for the two Fano lineshapes as function of temperature. 

To fit the temperature dependence of the pair of four step features seen in Fig. 4c (four on each side of zero bias), we use a series of symmetric Fermi-Dirac distribution functions as function of energy, $\epsilon$:

\begin{equation*}
\begin{split}
f_\mathrm{step}(\epsilon) & = \sum\limits_{i=1}^{4} \left(f_{FD}^+ + f_{FD}^-\right) + s\epsilon + c_2 \\
& = \sum\limits_{i=1}^{4} \left(a^+_i \frac{1}{1+e^{\frac{\epsilon+\epsilon_i}{k_{B}T}}} + a^-_i\left(1-\frac{1}{1+e^{\frac{\epsilon-\epsilon_i}{k_{B}T}}}\right)\right) + s\epsilon + c_2 
\end{split}
\end{equation*}

where \textit{i} is the step number, $a^+_i$ is the amplitude of the $i^{th}$ step for $\epsilon>0$, $a^-_i$ is the amplitude of the corresponding step at $\epsilon<0$, $\epsilon_i$ is the position of the $i^{th}$step, $k_B$ is the Boltzmann constant, $T$ is the temperature, s is the slope, and c$_2$ is a constant background term.

The recorded spectra are broadened by thermal contribution as well as the applied lock-in voltage. These effects have to be deconvoluted to get the intrinsic line-shape. To correct for the lock-in modulation voltage (V$_m$) we use the broadening function: 

\begin{equation*}
f_{V_{m}}(\epsilon) = \frac{2}{\pi}\Re\frac{\sqrt{V_m^2 - \epsilon^2}}{V_m^2}
\end{equation*}

where $\Re$ is the real part of a complex number. To account for thermal broadening due to
the temperature (T) of the tip, we use the derivative of the Fermi-Dirac distribution:

\begin{equation*}
f_{T}(\epsilon) = \frac{\partial}{\partial\epsilon}\left(\frac{1}{1+e^{\frac{\epsilon}{k_{B}T}}}\right)
\end{equation*}

Finally, the simulated LDOS is obtained by convolving these functions either,

\begin{equation*}
f_\mathrm{total}^\mathrm{Fano}(\epsilon) = f_\mathrm{Fano} * f_{V_{m}} * f_{T}
\end{equation*}

or,

\begin{equation*}
f_\mathrm{total}^\mathrm{step}(\epsilon) = f_\mathrm{step} * f_{V_{m}} * f_{T}
\end{equation*} 

The simulated LDOS is fitted to the experimental d$I$/d$V$ spectra to obtain the intrinsic linewidth $\Gamma$ in the first case and the step-positions in the second case.

\textit{DFT calculations.} Density functional theory calculations are performed with the full potential, all-electron, numeric atom-centered orbital code FHI-AIMS.\cite{aims1,aims2,Xinguo/implem_full_author_list,Levchenko/etal:2015} We use the standard FHI-AIMS `light' pre-constructed basis sets of numeric atomic orbitals. Supercell calculations are performed with a $8\times4$ $\Gamma$-centered $\mathbf{k}$-point sampling. We use the Perdew-Burke-Ernzerhof (PBE) generalized gradient approximation to the exchange-correlation functional.\cite{pbe} Van der Waals interactions are included with the pair-wise Tkatchenko-Scheffler correction.\cite{ts_dispersion} Atomic forces are relaxed to less than $10^{-2}$ eV/\AA. Vibrations are calculated with the finite difference method. Electron-phonon coupling constants are based on the electronic friction approach.\cite{askerka_prl_2016,maurer_prb_2016} In pursuit of open materials science,\cite{Himanen2019} the DFT relaxed geometry of the monolayer is available in the NOvel MAterials Discovery (NOMAD) repository \cite{NOMAD}.

\begin{suppinfo}
Spectroscopy on oxygen-intercalated graphene on Ir(111); disordered islands and checker-board phases of CTC; assembly of single component F$_4$TCNQ and TTF molecules on G/O/Ir(111); electronic band structure of CTC with or without graphene; temperature-dependent spectra -- experiment and simulated; molecular vibrations of CTC; and assembly and spectroscopy of CTC grown on G/Ir(111).
\end{suppinfo}

\begin{acknowledgement}

We thank Jose Lado for discussions. This research made use of the Aalto Nanomicroscopy Center (Aalto NMC) facilities and was supported by the European Research Council (ERC-2017-AdG no.~788185 ``Artificial Designer Materials'') and Academy of Finland (Academy professor funding nos.~318995 and 320555,  postdoctoral researcher nos.~309975 and 316347). We gratefully acknowledge high performance computing resources from the Aalto Science-IT project and the CSC-IT Center for Science, Finland.
\end{acknowledgement}

\bibliography{achemso-demo}

\newpage
\begin{center}
\Large{\textbf{Supplementary Information: \\ 
Electronic Characterization of a Charge-Transfer  \\
Complex Monolayer on Graphene}}
\end{center}

\setcounter{page}{1}

\renewcommand\thefigure{S\arabic{figure}} 
\renewcommand\thepage{S\arabic{page}} 

\noindent
\newpage
\textbf{Oxygen-Intercalated Graphene on Ir(111)}

\vspace{9pt}

Figure S1 contains d$I$/d$V$ spectroscopy recorded on oxygen-intercalated graphene grown on Ir(111) (G/O/Ir(111)). The d$I$/d$V$ spectrum in Fig.~S1a shows a phonon gap of $\pm$80 mV which indicates phonon-mediated inelastic tunnelling process that is intrinsic to graphene layer. Fig.~S1b shows field emission resonance peaks where the position of the first peak points to the fact that the work function of intercalated graphene is higher than the neutral graphene by $\sim$0.7 eV.

\vspace{9pt}

\begin{figure}[h]
  \includegraphics[width=0.9\textwidth]{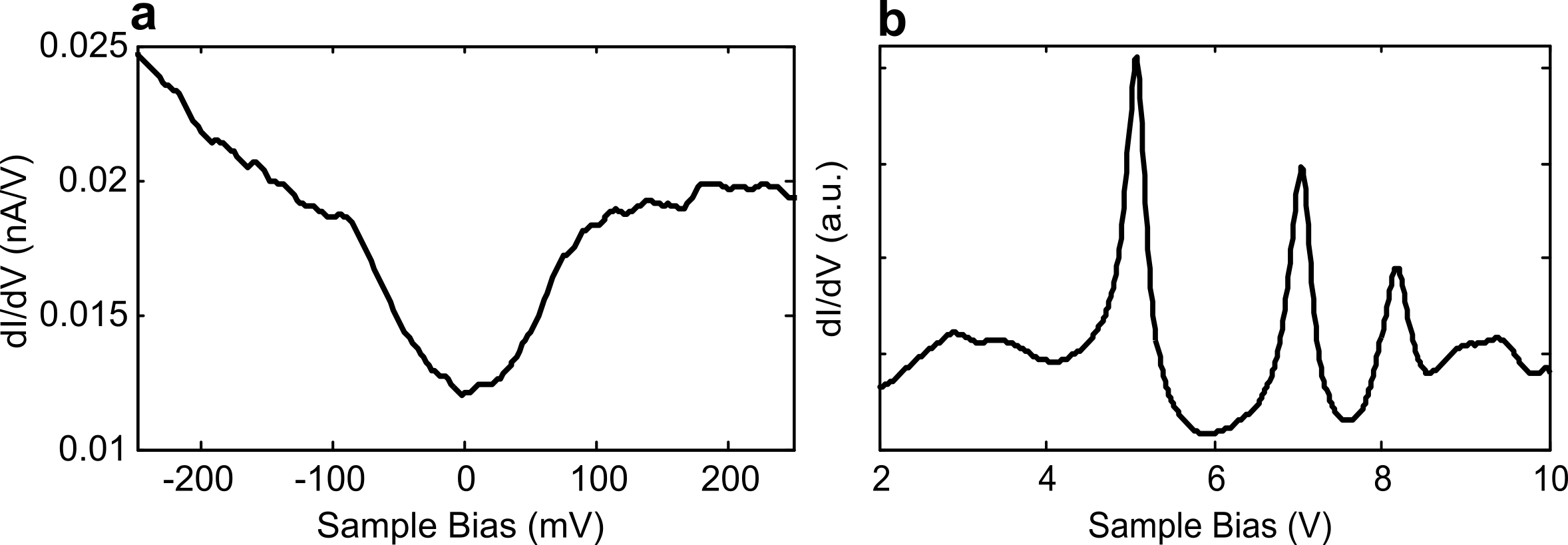}
  \caption{(a) Short range d$I$/d$V$ spectrum recorded on an oxygen-intercalated graphene on Ir(111) (G/O/Ir(111)) surface shows a phonon gap of $\pm$80 mV, which indicates that intercalated graphene is significantly decoupled \cite{Zhang2008}. The setpoint is 10 pA at 0.5 V. (b) Field emission resonance (FER) spectrum recorded on the surface has the first peak at $\sim$5.0 V. This points to a work function difference of $\sim$0.7 eV compared to neutral graphene. This enhancement is attributed to \emph{p}-type doping of graphene due to oxygen intercalation. FER spectrum was recorded at tunneling current of 0.5 nA.}
  \label{fig:af_dca}
\end{figure}

\newpage 

\noindent
\textbf{Disordered Islands of CTC}

\vspace{9pt}

Figure S2 shows an (a) overview and (b) zoomed-in STM topography images of disordered islands of CTC grown on G/O/Ir(111) surface. Disordered islands are result of sequential deposition of $\sim$0.25 monolayer of F$_4$TCNQ and TTF molecules on a G/O/Ir(111) surface at low substrate temperature ($\approx$100 K).  Annealing of such samples at room temperature for 15-45 minutes leads to the formation of highly ordered CTC islands.

\vspace{9pt}

\begin{figure}[h]
  \includegraphics[width=0.8\textwidth]{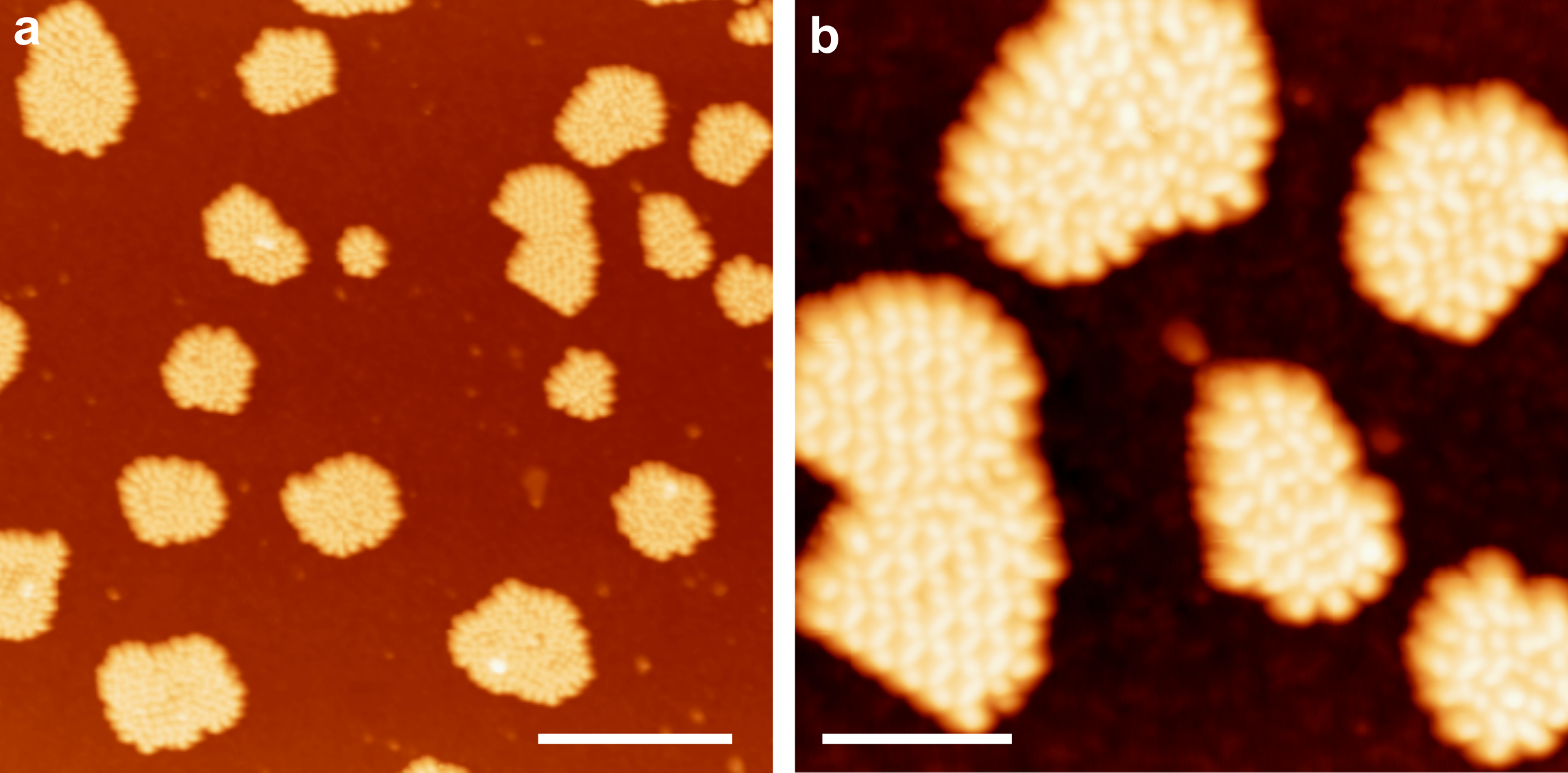}
  \caption{(a) An STM image of disordered islands of CTC on a G/O/Ir(111) surface. Imaging parameters: 0.8 pA and 1 V. The scale bar is 30 nm. (b) A zoomed-in STM image of a small area from panel (a) shows the arrangement of molecules in partially ordered and disordered islands. The bright features in the island correspond to either one or two TTF molecules. Imaging parameters: 0.8 pA and 1 V. The scale bar is 10 nm.}
  \label{fig:af_dca}
\end{figure}

\newpage

\noindent
\textbf{Checkerboard Phase of CTC}

\vspace{9pt}

Figure S3 shows STM topography image of checkerboard phase of the CTC with stoichiometry (F$_4$TCNQ)$_{x}$(TTF)$_{(1-x)}$. Between two checkerboard domains, we observe regular striped domains with rows of TTF and F$_4$TCNQ molecules as indicated by the rectangular box. 

\vspace{9pt}

\begin{figure}[h]
  \includegraphics[width=0.8\textwidth]{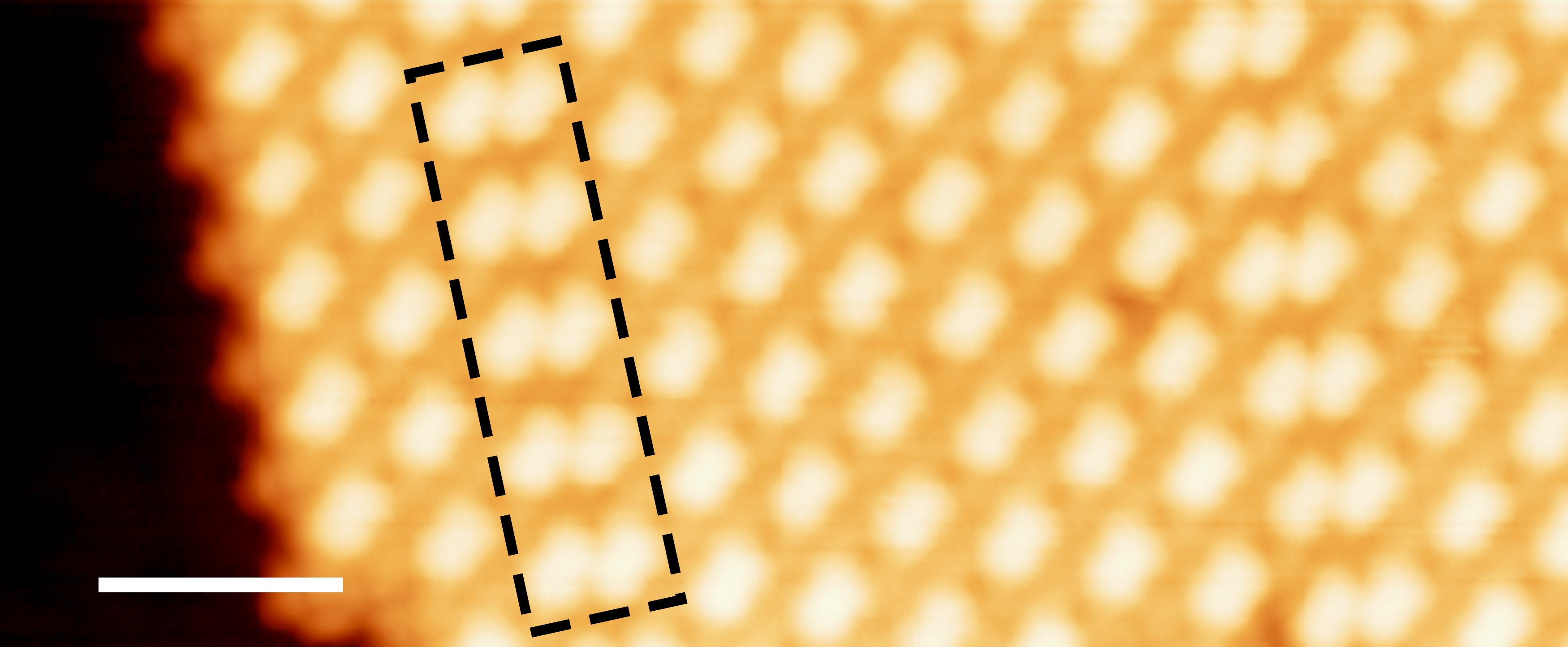}
  \caption{An STM topography image of the checkerboard phase of the CTC on a G/Ir(111) surface. The box indicates the boundary of two domains of the checkerboard phase. The boundary resembles the most common phase with alternate rows of TTF and F$_4$TCNQ molecules. Imaging parameters: 0.9 pA and 1.4 V. The scale bar is 3 nm.}
  \label{fig:checker_phase_topo}
\end{figure}

\newpage

\noindent
\textbf{Assembly of F$_4$TCNQ Molecules on G/O/Ir(111)}

\vspace{9pt}

Figure S4a shows that single-component F$_4$TCNQ molecules on G/O/Ir(111) surface assemble into chains. F$_4$TCNQ molecules of the chain remain neutral. Figure S4b shows DFT simulated LUMO of F$_4$TCNQ molcule. 

\vspace{9pt}

\begin{figure}[h]
  \includegraphics[width=0.6\textwidth]{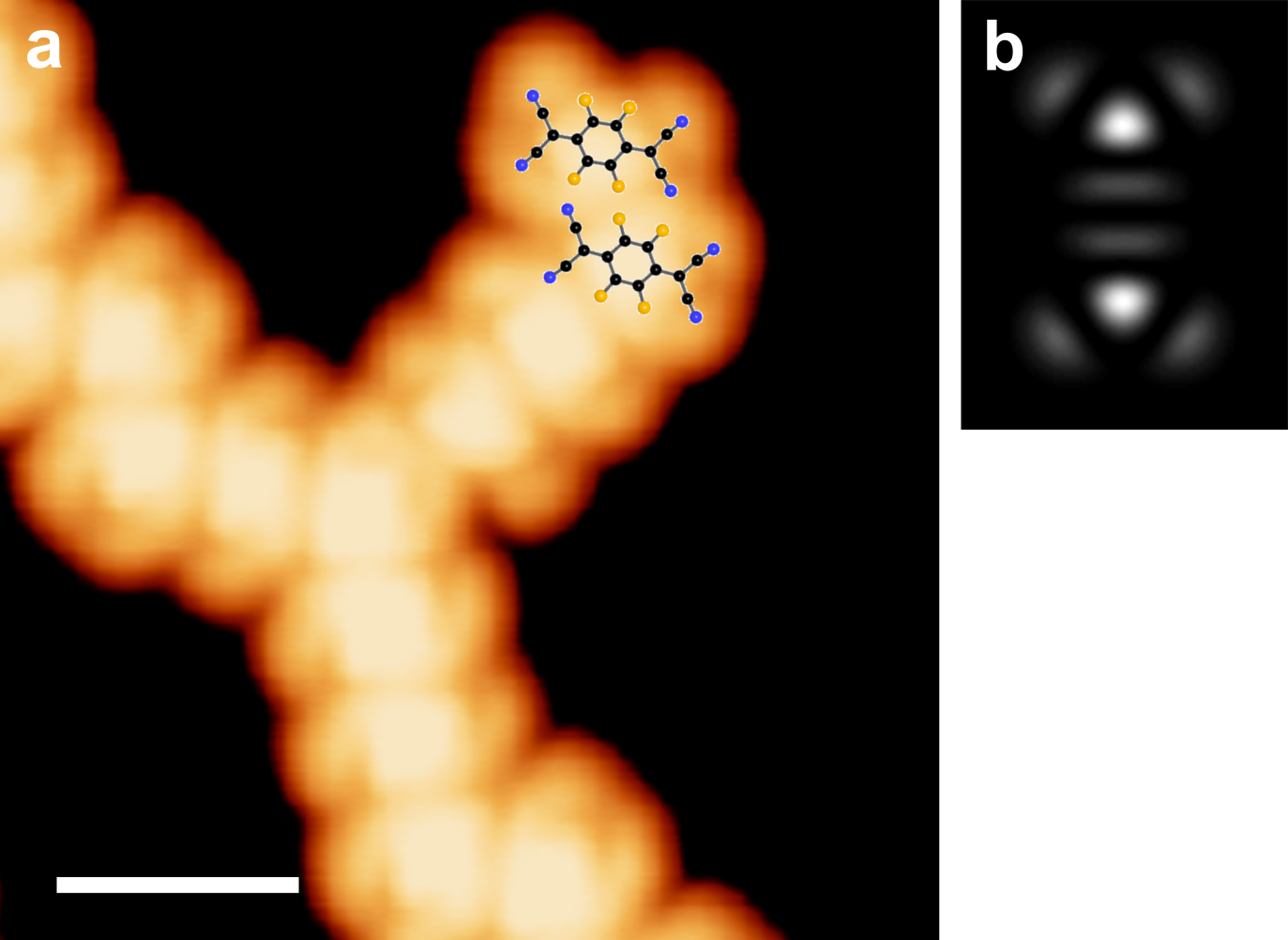}
  \caption{(a) Zig-zag arrangement of F$_4$TCNQ in F$_4$TCNQ molecular chains on a G/O/Ir(111) surface. Molecular structure of F$_4$TCNQ has been overlaid for clarity. Imaging parameters: 2 pA and 1.7 V. Scale bar is 2 nm. (b) A DFT simulated lowest unoccupied molecular orbital (LUMO) of F$_4$TCNQ.}
  \label{fig:af_dca}
\end{figure}

\newpage

\noindent
\textbf{Assembly of TTF Molecules on G/O/Ir(111)}

\vspace{9pt}

Figure S5a shows that the single-component TTF molecules on G/O/Ir(111) assemble into a close-packed island. Similar to single-component F$_4$TCNQ chains on the surface, TTF molecules in the island are also neutral. Figure S5b shows DFT simulated HOMO of TTF molecule. 

\vspace{9pt}

\begin{figure}[h]
  \includegraphics[width=0.6\textwidth]{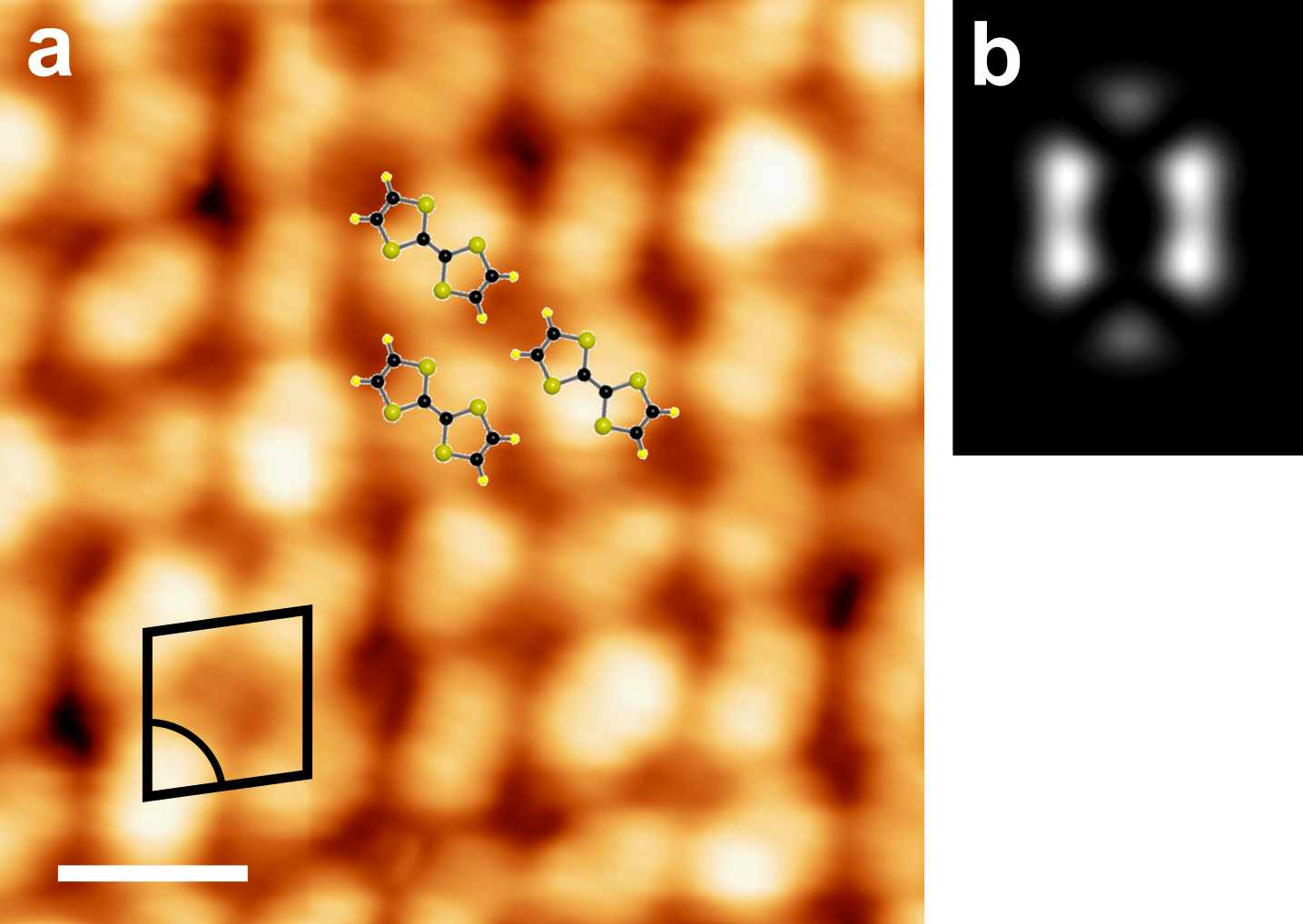}
  \caption{(a) An STM image of a small TTF island shows an oblique unit cell of the TTF assembly with unit cell of 8.0 $\pm$ 0.2 \r{A} and angle 83$^\circ$. The unit cell and molecular structure of TTF have been overlaid for clarity. Imaging parameters: 0.7 pA and 0.6 V. Scale bar is 1 nm. (b) A DFT simulated highest occupied molecular orbital (HOMO) of TTF.}
  \label{fig:af_dca}
\end{figure}

\newpage

\noindent
\textbf{Electronic Band Structure of CTC With or Without Graphene}

\vspace{9pt}

Figure S6a shows band structure of CTC calculated using DFT with and without graphene for spin-up configuration. The band structure of the spin-down configuration (not shown) looks similar to that of the spin-up configuration. Its apparent from the band structure that electronic coupling is asymmetric in various lattice directions and the maximum width of the dispersive band is $\sim$150 meV. This indicates that the lateral overlap of the molecular orbitals in the CTC is relatively weak. The right panel shows total DOS of the CTC. Figure S6b is the zoomed-in plot of the same band structure. 

\vspace{9pt}

\begin{figure}[h]
  \includegraphics[width=0.59\textwidth]{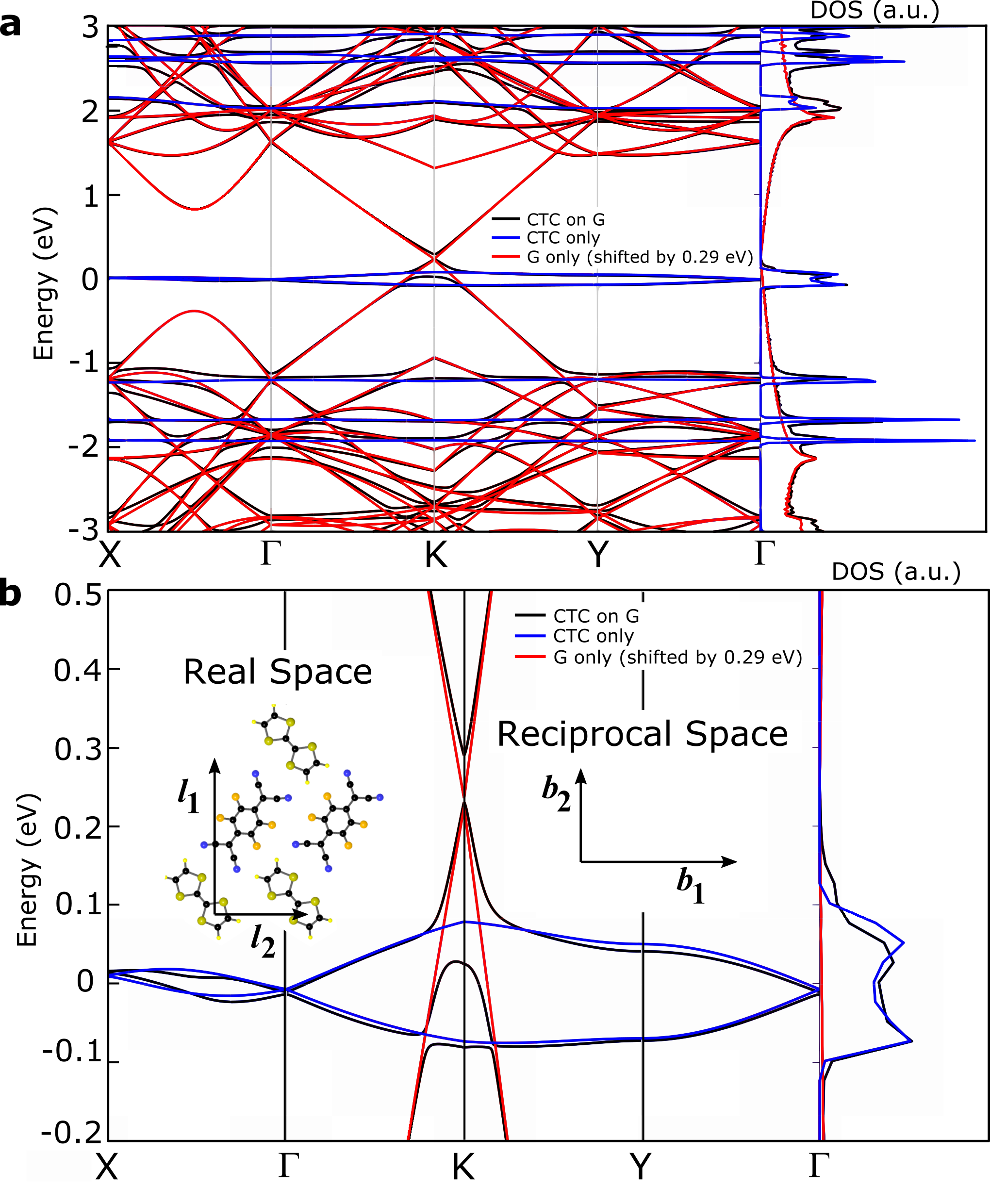}
  \caption{(a) Electronic band structure (left panel) and DOS (right panel) of CTC for its spin-up configuration. The band structure in the spin-down configuration looks very similar to that of the spin-up configuration. (b) A zoomed-in plot of the band structure from panel (a). 
  The left and right insets show the real and the reciprocal space unit cell, respectively. The rectangular unit cell has been chosen for easier understanding of band structure in reciprocal space. Here, X = $b_1$/2 and Y = $b_2$/2.
}
  \label{fig:bandstruct}
\end{figure}

\newpage

\noindent
\textbf{Temperature-Dependent Spectra -- Experiment and Simulated}

\vspace{9pt}

Figure S7a contains temperature-dependent d$I$/d$V$ spectra recorded on F$_4$TCNQ site in the CTC. The spectra has two types of features: an asymmetric dip at the Fermi energy and steps at energies $\pm$31 meV, $\pm$35 meV, $\pm$52 meV. Further, the asymmetric dip can be deconvoluted into a symmetric dip and a step at the Fermi-energy. As the temperature increase the amplitude of the dip reduces and vanishes at 20 K. However, the step remains visible at temperature 20 K. 
Panel b shows simulated temperature dependence of d$I$/d$V$ spectra assuming the the central dip has its origin due to IETS. We observe that the central dip doesn't vanish even at the temperature 20 K ruling out IETS origin of the dip. 

\vspace{9pt}

\begin{figure}[h]
  \includegraphics[width=0.75\textwidth]{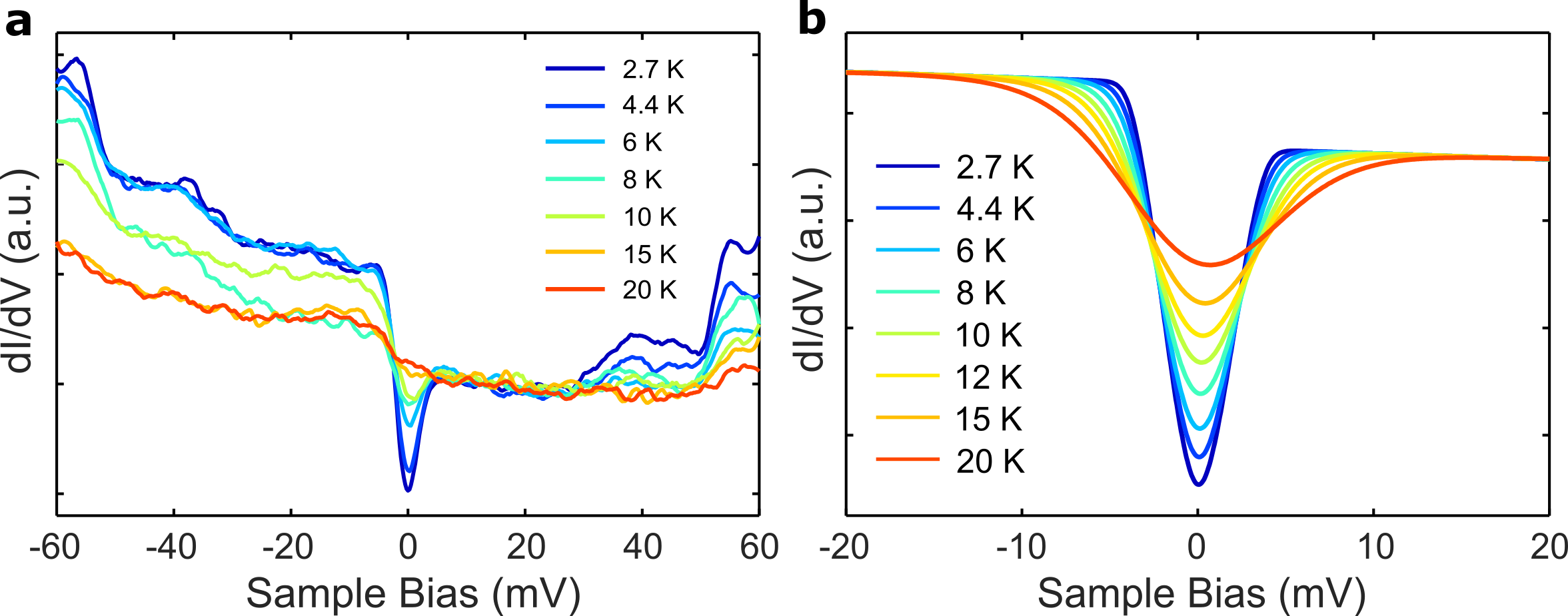}
  \caption{(a) Temperature-dependent d$I$/d$V$ spectra recorded on a F$_4$TCNQ site in the CTC. Similar to the TTF site, the dip vanishes with increasing temperature with relatively weak asymmetry still present at 20 K. (b) Simulated temperature dependence of the d$I$/d$V$ spectra assuming that the central dip would correspond to inelastic steps, where we have used the parameters extracted from the fitting of the IETS spectrum at 2.7 K. Further, each spectrum has been subjected to thermal and modulation broadening using V$_{amp}$ = 2 mV for all spectra.}
  \label{fig:af_dca}
\end{figure}

\newpage

\noindent
\textbf{Molecular Vibrations of CTC}

\vspace{9pt}

Figure S8 shows total DOS of the calculated modes of the CTC as function of energy and the illustration of two vibrational modes corresponding to 30 meV and 53 meV. 

\vspace{9pt}

\begin{figure}[h]
  \includegraphics[width=0.9\textwidth]{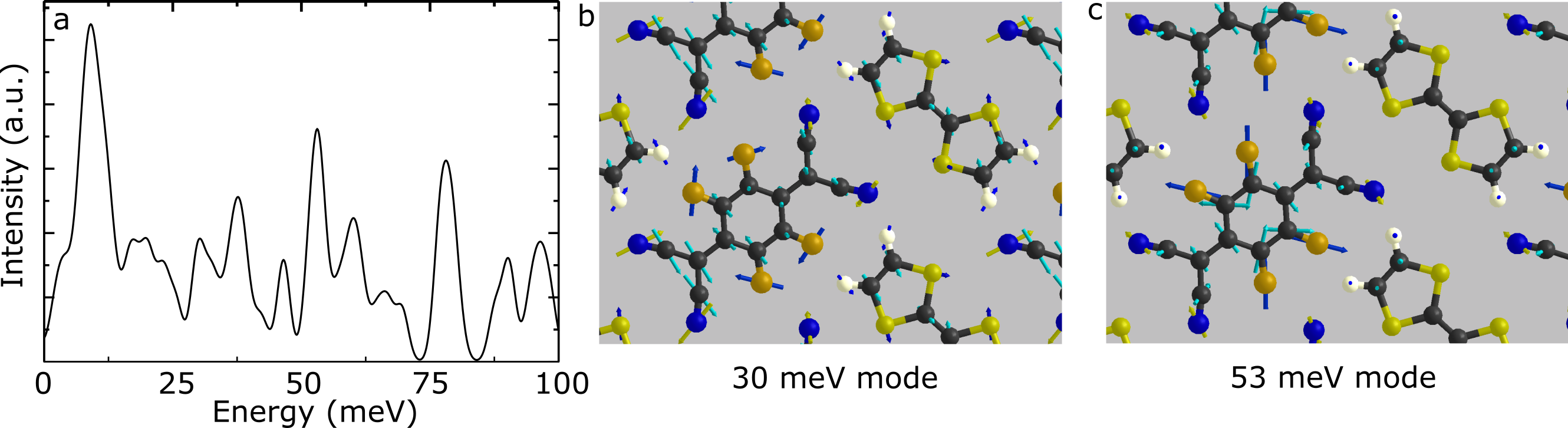}
  \caption{(a) Total DOS of the calculated vibrational modes of CTC as function of energy. (b) Vibrations in the CTC corresponding to the vibrational energy mode of 30 meV. (c) Vibrations in the CTC corresponding to the vibrational energy mode of 53 meV. }
  \label{fig:vibrations}
\end{figure}

\newpage

\noindent
\textbf{Assembly of CTC on G/Ir(111)}

\vspace{9pt}

Figure S9 shows CTC assembly on G/Ir(111) (without oxygen intercalation) and the short-range  d$I$/d$V$ spectrum recorded on TTF molecule of the CTC. Here, we also observe a dip at the Fermi-energy of a comparable width as that on CTC grown on G/O/Ir(111) surface. 

\vspace{9pt}

\begin{figure}[h]
  \includegraphics[width=0.9\textwidth]{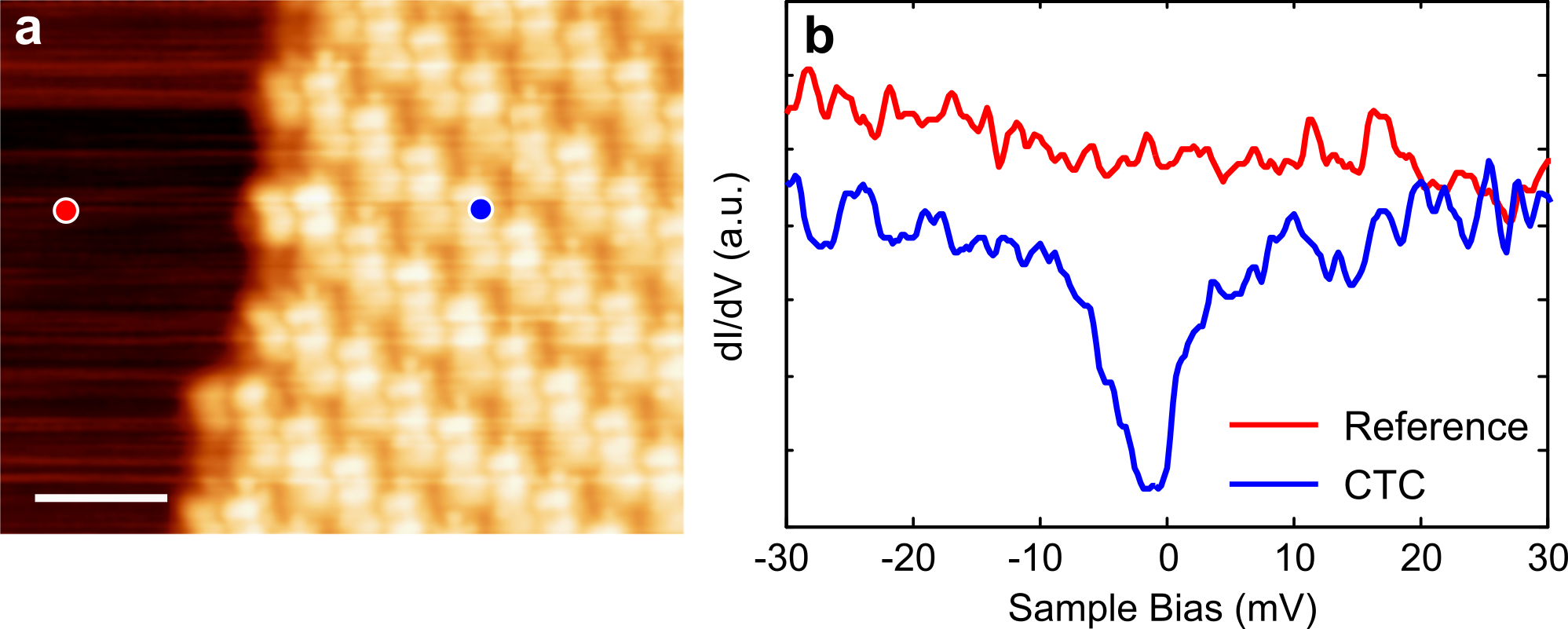}
  \caption{(a) An STM image showing a CTC assembly on graphene on a Ir(111) (G/Ir(111)) surface. Similar to that on the G/O/Ir(111) surface, the CTC has alternate rows of TTF and F$_4$TCNQ molecules. Imaging parameters: 10 pA and 0.1 V. Scale bar is 2 nm. (b) Short range d$I$/d$V$ spectrum recorded on a TTF molecule in the CTC island shows a dip at zero sample bias. The width of the zero-bias dip is comparable to that of a CTC on G/O/Ir(111) surface. A spectrum recorded on a G/Ir(111) surface is presented as a reference.}
  \label{fig:af_dca}
\end{figure}

\end{document}